\documentclass[11pt,amsmath,amssymb,latexsym,a4paper]{article}
\usepackage{graphicx}  
\usepackage{dcolumn}   
\usepackage{bm}        
\usepackage{verbatim}  

 \usepackage{color} 
 
\newtheorem{myDefinition}{Definition}

\newtheorem{myProperty}{Property}

\newcommand{\ket}[1]{\left|{#1}\right>}

\newcommand{\bra}[1]{\left<{#1}\right|}

\newcommand{\toshIdentity}{{\mathbf{1}}}

\newcommand{\highlight}[1] {#1}

\begin{document}

\title{\highlight{An alternative} to test experimentally Everett's theory}
\author{Hitoshi Inamori\\
}

\bigskip

\date{\today}

\maketitle

\begin{abstract}
It is generally accepted that Everett's theory of quantum mechanics cannot be experimentally tested as such experiment would involve operations on the observer which are beyond our current technology. We propose an alternative to test Everett's theory which does not involve any operation on the observer. If we assume that the observer is of finite dimension, it is shown that Everett's theory leads to distinctive properties for the system being observed, and that such difference can be experimentally tested.

\bigskip


\end{abstract}

\section{Introduction}  

According to quantum theory, performing a measurement on a physical system leads in general to a truly random outcome which cannot be predicted. Many theories or interpretations of quantum physics try to explain how this random outcome arises. 

These theories usually agree on how the measurement starts~\cite{vonNeumann}: when an observer performs a measurement on a system, the two physical systems interact so that the state of the combined system becomes entangled. It becomes a sum of states, and in each of these states, the observer witnesses an outcome and the observed system is in the state corresponding to this outcome.

The theories however diverge from this point. In particular, the Copenhagen theory~\cite{Copenhagen} asserts that the state of the combined system ``jumps'' from the above superposition in which the outcome is not \highlight{determined}, to the final state, in which only one of the possible outcomes occurs. This irreversible process is often referred to as the ``collapse of the wave function'' where ``wave function'' is synonym for the state of the system. 

On the other hand, Everett's theory~\cite{Everett} asserts that such irreversible process does not occur. There are many variants for Everett's theory but in this paper we shall consider the original ``Relative State'' version~\cite{Everett}, which asserts that the combined system remains in a superposition of states in each of which one version of the observer actually witnesses one possible outcome of the measurement. 

\highlight{In this paper we will choose to} distinguish the above two theories as follows:

\begin{myDefinition} In the Copenhagen theory, when an observer performs a measurement, the system state jumps irreversibly from a superposition of states to a single state in which only one outcome occurs.

In Everett's theory, such wave function collapse does not occur. The observed system and the observer remain in an entangled superposition of states.
\end{myDefinition}

It is believed that in practice the two theories lead to the same predictions that can be tested experimentally. However, several thought experiments have been proposed~\cite{Deutsch, Lockwood, Vaidman} to test Everett's theory against the Copenhagen one. In particular, Deutsch~\cite{Deutsch} proposed that if the above interaction between the observer and the observed system could be reversed, then we could experimentally have the observer perform a measurement on a system, have him declare that the measurement was done without him revealing the outcome of the measurement, and then reverse the interaction between the observer and the observed system. If the initial state for the observed system is in a superposition of states, then according to Everett's theory the observed system should remain in the same superposition of states after this entire process. In contrast, according to the Copenhagen theory the observed system would jump into one of the states corresponding to a possible measurement outcome. Such discrepancy could be tested experimentally. However, Deutsch's experiment would require that we can revert the interaction process involving the observer, which is assumed to be a relatively large system.  As such it remains a Gedanken experiment, not realizable experimentally.

In this paper we propose an alternative to test Everett's theory that does not require any transformation to be made on the observer after the measurement. The key observation is that an observer is a physical system which is finite, and that this finite dimensionality implies a certain constraint on the state of the observed system after its interaction with the observer. 

More precisely, suppose that we have a very long sequence of qubits, all prepared in the state $\frac{\ket{0}+\ket{1}}{\sqrt{2}}$. Assume that we have an observer performing a measurement on each of these qubits in the $\{\ket{0},\ket{1}\}$ basis. If the observer had an infinite memory then he could remember the result of all the measurements and the observed system would be in a fully mixed state, corresponding to a stream of qubits in which each qubit is in the state $\ket{0}$ (corresponding to the case in which the observer witnessed $\ket{0}$) or in the state $\ket{1}$ (corresponding to the case in which the observer witnessed $\ket{1}$) with equal classical probability 50\%.     

Now if the observer is finite, then his memory is finite as well, and intuitively we can feel that the observer cannot ``fully entangle'' with all the qubits of this very long sequence. Now independently of the size of the memory, according to the Copenhagen theory, the observer would collapse the wave function of the system, and the very long sequence of the qubits would be in a fully probabilistic mixture of $\ket{0}$ and $\ket{1}$'s. The purpose of this paper is to show that with Everett's theory the state of the observed system is provably different from what the Copenhagen theory predicts. More precisely, we give a sufficient condition on the length of the sequence of the qubits compared to the observer's dimensionality so that Everett's theory gives different experimental predictions compared to the Copenhagen theory.

Such experiment \highlight{may be} easier to perform than Deustch's Gedanken experiment described above. In particular we do not need to know the inner working of the observer as we do not need to revert any process involving the observer. The only requirement is that we can perform measurement on the observed system in different bases so that one of such bases leads to outcome statistics different from what is predicted by the Copenhagen theory.

\section{Main result}

The following constitutes the main result of this paper:

\begin{myProperty}
Let $N$ and $m$ be positive integers, and let a source of qubits emit $N\times m$ qubits, each of them in the state $(\ket{0}+\ket{1})/\sqrt{2}$. The source is therefore emitting $m$ streams of $N$ qubits which are all in the same initial state. Let us ask the observer $O$ to perform a measurement on each of the $N\times m$ qubits in the basis $\{\ket{0},\ket{1}\}$ and let us ask the observer to send through the observed qubits.  \highlight{We suppose that the observer interacts with each qubit one at a time, although it is sufficient to assume that the observer interacts with one N-qubit stream at a time.}

Let $\rho_1,\rho_2,\ldots, \rho_m$ represent the reduced density matrix of the $m$ streams, where for each $i=1,\ldots,m$, $\rho_i$ is a $2^N$ dimensional density matrix obtained from the density matrix representing the entire experimental setup by tracing out the space complementary to the Hilbert space for the $i$-th stream.

We want to consider a density matrix representing each stream of $N$-qubits, which we obtain by shuffling the $m$ streams in a random order. In this case, the reduced density matrix is identical for all the streams and is equal to the average of the reduced density matrices for the $m$ streams of qubits before the shuffling. We denote by $\rho_S = \frac{1}{m}(\rho_1+\cdots+\rho_m)$ this reduced density matrix. Then, 

\begin{itemize}
\item according to the Copenhagen theory, the reduced density matrix $\rho_S$ is fully mixed, i.e. $\rho_S =\frac{1}{2^N}\toshIdentity_{2^N}$.
\item according to Everett's theory, if the dimension of the observer is finite then for a sufficiently large $N$, the reduced density matrix $\rho_S$ is not fully mixed. 
\end{itemize}

As we have $m$ instances of the $N$-qubit streams having the same reduced density matrix $\rho_S$, one can experimentally distinguish whether the density matrix $\rho_S$ is in accordance with the Copenhagen theory or Everett's theory.
\label{mainResult}
\end{myProperty}

The first part of the property giving the reduced density matrix $\rho_S$ according to the Copenhagen theory is straightforward to prove. Indeed, the observer is measuring each qubit in the basis $\{\ket{0},\ket{1}\}$ and the initial state of the qubit is $(\ket{0}+\ket{1})/\sqrt{2}$. According to the Copenhagen theory, each qubit jumps to the state $\ket{0}$ or $\ket{1}$ with probability $\frac{1}{2}$ when the measurement is performed. As a result, the density matrix for any stream of $N$ qubits is $\frac{1}{2^N}\toshIdentity_{2^N}$ after the measurement performed by the observer.

Proving the second part of the property is less straightforward and is done in the following two sections.

\section{Modeling the experiment}

In this section we give a mathematical description of the experiment and in particular the description of the observer as a quantum system interacting with each stream of $N$ qubits. This description will allow us to precise what we mean by the dimensionality of the observer. It will later also allow us to give a constraint on the dimensionality of the reduced density state for the $N$-qubit systems after interaction with the observer.

Let $H_O$ be the Hilbert space describing the observer and let $H_S$ be the Hilbert space describing one stream of $N$ qubits. Without loss of generality suppose that the observer is initially in a state $\ket{\phi_0}\in H_O$. As we have seen, all the $m$ streams of $N$ qubits are initially in the state $(\ket{0}+\ket{1})^N/\sqrt{2^N}$. \highlight{We have assumed that the observer interacts with one $N$-qubit stream at a time:} the observer performs a measurement on the first stream of $N$ qubits, and its interaction can be represented by an unitary operator $U(1)$ on $H_S\otimes H_O$ where $H_S$ is the Hilbert space for the first $N$-qubit stream. Then the observer performs a measurement on the second stream, and the interaction is represented by an unitary operator $U(2)$ on $H_S\otimes H_O$ where $H_S$ is now the Hilbert space for the second stream, etc. \highlight{(we could explicitly perform a swap of the $k$-th stream of qubits with an auxiliary system of $N$ qubits and perform the unitary operator on this auxiliary system combined with the observer, swapping back the $k$-th stream after the measurement is done. This would lead to an equivalent setup.)} A priori, there is no reason why $U(1)$ should be identical to $U(2)$ as we do not know the inner working of the observer, except the fact that it is of finite dimension (Figure~\ref{Figure1}).

\setlength{\unitlength}{0.025 in}
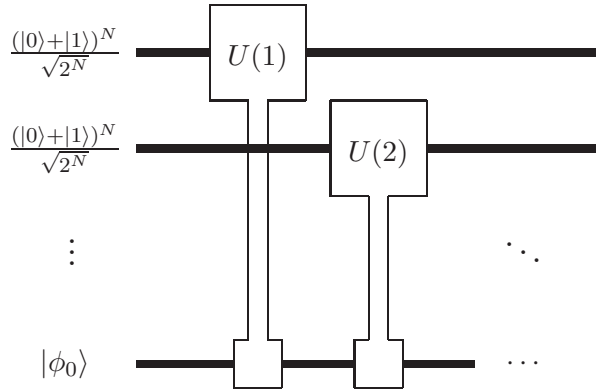
\begin{figure}[!h]
\centering
\begin{picture}(100,100)

\newsavebox{\upperbox}
\savebox{\upperbox}(20,20){
\thinlines
\put(0,20){\line(1,0){20}}
\put(0,0){\line(1,0){8}}
\put(12,0){\line(1,0){8}}
\put(0,0){\line(0,1){20}}
\put(20,0){\line(0,1){20}}
}

\newsavebox{\lowerbox}
\savebox{\lowerbox}(20,10){
\thinlines
\put(5,0){\line(1,0){10}}
\put(5,10){\line(1,0){3}}
\put(12,10){\line(1,0){3}}
\put(5,0){\line(0,1){10}}
\put(15,0){\line(0,1){10}}
}

\put(0,80){\makebox(0,0){$\frac{(\ket{0}+\ket{1})^N}{\sqrt{2^N}}$}}
\linethickness{1mm}
\put(15,80){\line(1,0){15}}
\put(50,80){\line(1,0){60}}
\put(20,70){\usebox{\upperbox}}
\put(40,79){\makebox(0,0){$U(1)$}}

\put(0,15){\makebox(0,0){$\ket{\phi_0}$}}
\linethickness{1mm}
\put(15,15){\line(1,0){20}}
\put(20,10){\usebox{\lowerbox}}

\thinlines
\put(38,20){\line(0,1){50}}
\put(42,20){\line(0,1){50}}

\put(0,60){\makebox(0,0){$\frac{(\ket{0}+\ket{1})^N}{\sqrt{2^N}}$}}
\linethickness{1mm}
\put(15,60){\line(1,0){40}}
\put(75,60){\line(1,0){35}}
\put(45,50){\usebox{\upperbox}}
\put(65,59){\makebox(0,0){$U(2)$}}

\linethickness{1mm}
\put(45,15){\line(1,0){15}}
\put(45,10){\usebox{\lowerbox}}

\thinlines
\put(63,20){\line(0,1){30}}
\put(67,20){\line(0,1){30}}

\put(0,30){\makebox(3,20){$\Large\vdots$}}

\put(90,35){\makebox(10,10){$\Large\ddots$}}

\linethickness{1mm}
\put(70,15){\line(1,0){15}}
\put(90,10){\makebox(10,10){$\Large\cdots$}}

\end{picture}
\caption{Model for the experimental setup}\label{Figure1}
\end{figure}

Note however that a time dependent operator $U(i)$ is equivalent to a time independent operator coupled with an additional physical system, playing the role of a ``clock'' internal to the observer. In other words, the time dependent $U(i)$ can be replaced by a time independent operator $U$ applying $U(i)$ depending on an additional variable $i$ on the observer side, which is incremented each time a measurement is performed. The internal variable $i$ can be represented by $\lceil\log_2 m\rceil$ qubits. For instance a time dependent operator $U(i)$ for $i=1, 2, 3, 4$ can be replaced by a constant operator $U$ and 2 additional qubits for the observer's state as in Figure~\ref{Figure2}. \highlight{We will assume henceforth that the additional qubits to encode the internal variable $i$ is part of the observer $O$.}

\setlength{\unitlength}{0.025 in}
\begin{figure}[!h]
\centering
\begin{picture}(190,70)

\newsavebox{\operator}
\savebox{\operator}(20,60){
\thinlines
\put(0,40){\line(1,0){20}}
\put(0,20){\line(1,0){8}}
\put(12,20){\line(1,0){8}}
\put(0,20){\line(0,1){20}}
\put(20,20){\line(0,1){20}}

\put(5,0){\line(1,0){10}}
\put(5,10){\line(1,0){3}}
\put(12,10){\line(1,0){3}}
\put(5,0){\line(0,1){10}}
\put(15,0){\line(0,1){10}}

\put(8,10){\line(0,1){10}}
\put(12,10){\line(0,1){10}}
}

\linethickness{1mm}
\put(5,60){\makebox(0,0){$H_S$}}
\put(5,35){\makebox(0,0){$H_O$}}
\put(15,60){\line(1,0){15}}
\put(15,35){\line(1,0){20}}
\put(20,20){\usebox{\operator}}
\put(40,59){\makebox(0,0){$U(1)$}}
\put(50,60){\line(1,0){10}}
\put(45,35){\line(1,0){20}}
\put(50,20){\usebox{\operator}}
\put(70,59){\makebox(0,0){$U(2)$}}
\put(80,60){\line(1,0){10}}
\put(75,35){\line(1,0){20}}
\put(80,20){\usebox{\operator}}
\put(100,59){\makebox(0,0){$U(3)$}}
\put(110,60){\line(1,0){10}}
\put(105,35){\line(1,0){20}}
\put(110,20){\usebox{\operator}}
\put(130,59){\makebox(0,0){$U(4)$}}
\put(140,60){\line(1,0){48}}
\put(135,35){\line(1,0){53}}

\thinlines
\put(40,30){\line(0,-1){8}}
\put(40,18){\line(0,-1){6}}
\put(40,20){\circle{4}}
\put(40,10){\circle{4}}

\put(70,30){\line(0,-1){8}}
\put(70,18){\line(0,-1){6}}
\put(70,20){\circle*{4}}
\put(70,10){\circle{4}}

\put(100,30){\line(0,-1){8}}
\put(100,18){\line(0,-1){6}}
\put(100,20){\circle{4}}
\put(100,10){\circle*{4}}

\put(130,30){\line(0,-1){8}}
\put(130,18){\line(0,-1){6}}
\put(130,20){\circle*{4}}
\put(130,10){\circle*{4}}

\put(15,20){\line(1,0){23}}
\put(42,20){\line(1,0){26}}
\put(72,20){\line(1,0){26}}
\put(102,20){\line(1,0){26}}
\put(132,20){\line(1,0){56}}

\put(15,10){\line(1,0){23}}
\put(42,10){\line(1,0){26}}
\put(72,10){\line(1,0){26}}
\put(102,10){\line(1,0){26}}
\put(132,10){\line(1,0){56}}

\put(158,20){\circle*{4}}
\put(158,20){\line(0,-1){13}}
\put(158,10){\circle{6}}

\put(173,20){\circle{6}}
\put(173,23){\line(0,-1){6}}

\put(5,10){\makebox(0,0){$\ket{0}$}}
\put(5,20){\makebox(0,0){$\ket{0}$}}

\linethickness{1mm}

\end{picture}
\caption{A time dependent operator $U(i)$, $i=1,2,3,4$ can be replaced by a constant operator and $\log_2 4 = 2$ additional qubits. The operator $U(i)$ is applied only when each auxiliary qubit is $\ket{0}$ if the circle is white or $\ket{1}$ if the circle is black. The $\oplus$ represents the NOT operator. The last two gates on the auxiliary qubits increment the counter formed by the two auxiliary qubits.}\label{Figure2}
\end{figure}
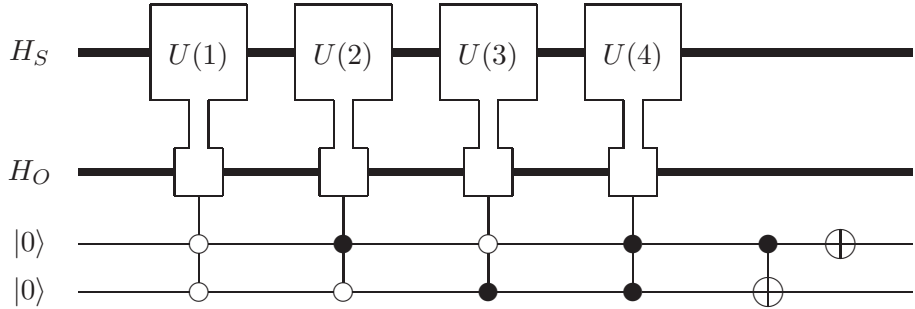

We have therefore shown that the experiment can be described by a successive interaction of the observer $O$ with the $N$-qubit streams $i=1,\ldots,m$ and that the interaction can be represented by an identical unitary operator $U$ acting on $H_S\otimes H_O$ where $H_S$ is the Hilbert space corresponding to one $N$-qubit stream.

\section{Demonstration of the main result}

We are now ready to prove Property~\ref{mainResult}. We denote by $D$ the dimension of the Hilbert space $H_O$ describing the observer. By hypothesis of the Proposition~\ref{mainResult}, $D$ is finite. Suppose that the size of the streams $N$ is so that $2^N$ is strictly larger than $D^2$. Then we can show the following property:

\begin{myProperty} There exists a subspace $F_S$ of $H_S$ of dimension less than $D^2$ such that whichever the initial state for the observer, the density matrix $\rho_S$ for the observed $N$ qubit stream has its support in the subspace $F_S$. In particular, for any state $\ket{\chi}$ which is orthogonal to $F_S$, we have $\bra{\chi}\rho_S\ket{\chi} = 0$. 
\end{myProperty}

To prove this property we first remind the Schmidt decomposition theorem (the reader may find a proof for instance in~\cite{Bergou}):

\begin{myProperty} Let $H=H_A\otimes H_B$ be the product space of Hilbert spaces $H_A$ and $H_B$. Then for any state $\ket{\psi_{AB}}\in H$, there exist orthonormal bases $\{\ket{a_i}\}_i$ for $H_A$ and $\{\ket{b_i}\}_i$ for $H_B$ such that:

\begin{equation}
\ket{\psi_{AB}} = \sum_{k=1}^n \alpha_k \ket{a_k}\otimes\ket{b_k}
\end{equation}
where $n \leq \min (\dim(H_A),\dim(H_B))$ and $\alpha_k$ are non-negative real numbers.
\end{myProperty}

Let $\{\ket{e_i}\}_{i=1,\ldots,D}$ be an orthonormal basis for $H_O$. If the observer's initial state is $\ket{e_i}$ then after the observer performs the measurement on an $N$-qubit system initially prepared in the state $(\ket{0}+\ket{1})^N/\sqrt{2^N}$, the combined system is in the state:

\begin{equation}
\frac{(\ket{0}+\ket{1})^N}{\sqrt{2^N}}\otimes\ket{e_i}\mapsto U\highlight{\left[\frac{(\ket{0}+\ket{1})^N}{\sqrt{2^N}}\otimes\ket{e_i}\right]}\in H_S\otimes H_O,
\end{equation}
which, according to Schmidt decomposition theorem, reads:
\begin{equation}
U\highlight{\left[\frac{(\ket{0}+\ket{1})^N}{\sqrt{2^N}}\otimes\ket{e_i}\right]} = \sum_{j=1}^D \alpha^i_j \ket{\psi^i_j}\otimes \ket{\phi^i_j}
\end{equation}
for some non-negative real numbers $\alpha^i_j$ (some of them possibly zero), and orthonormal bases $\{\ket{\psi^i_j}\}_j$ of $H_S$ and $\{\ket{\phi^i_j}\}_j$ of $H_O$ with $j=1,\ldots,D$. The superscript $i$ reminds us that these orthonormal bases may depend on the observer state $\ket{e}_i$ initially chosen.

Let $F_S$ be the subspace of $H_S$ spanned by the $D^2$ states $\{ \ket{\psi^i_j} \}_{i,j}$, $i, j=1,\ldots
,D$. The dimension of $F_S$ is therefore less than, or equal to, $D^2$. Now whichever the state $\ket{\phi}$ for the observer, it reads as a linear combination of the orthonormal states $\ket{e_i}$. Thus, after the observer performs the measurement on the $m$ streams of $N$-qubits, the state describing each stream of qubits lies within the subspace $F_S$, in the sense that, for each stream $k\in\{1,\ldots,m\}$, the state $\ket{\psi}$ of the entire experimental setup reads:

\begin{equation}
\ket{\psi} = \sum_{i,j}\ket{\eta^k_{i,j}}\otimes \ket{\psi^i_j}
\end{equation}
where the $\ket{\psi^i_j}$ are states defined above describing the $k$-th stream, and $\ket{\eta^k_{i,j}}$ are states describing the remaining $(m-1)$ streams and the observer (the states $\ket{\eta^k_{i,j}}$ are possibly not normalized nor orthogonal with each others).

We deduce that for any stream of $N$ observed qubits, the reduced density matrix $\rho_S$ describing the stream is such that

\begin{equation}
\bra{\chi}\rho_S\ket{\chi} = 0
\end{equation}
for any state $\ket{\chi}$ which is orthogonal to $F_S$. In other words, the density matrix $\rho_S$ has support in the subspace $F_S$, which completes the proof.

We have shown that in the absence of wave function reduction, the density matrix describing the stream of $N$ qubits have support in $F_S$ of dimension which is limited by $D^2$. If $D^2$ is lower than $2^N$ then the density matrix of the $N$ qubits cannot be fully mixed. In particular, for any state $\ket{\chi}$ in the subspace $F_S^\perp$ orthogonal to $F_S$ ($F_S^\perp$ has dimension larger than, or equal to, $2^N-D^2$), the probability to observe $\ket{\chi}$ from the stream of $N$ qubits is zero while the probability is $1/2^N$ with the fully mixed state. \highlight{Another example, if you take a random state $\ket{\psi}$ for the $N$-qubit system and measure whether the $N$-qubit systems are in this state then we find that the probability would be between 0 and $q$ with $q \geq 1/2^{(2d)}$ while you should have this probability equal to $1/2^N$ with the fully mixed state.}

We have shown that it is sufficient that $2^N$ is larger than $D^2$ so that a theory without wave function reduction, such as Everett's, leads to a density state for the observed $N$ qubits which is distinct from the fully mixed state. This completes the proof for Property~\ref{mainResult}.

Let us illustrate the above discussion with a toy example. Assume that the observer can be described by a single qubit, meaning $D=2$. Let us take $N=3$, thus $2^N = 8 > D^2$. The unitary operator $U$ representing the interaction between the streams of $N=3$ qubits and the observer can be of any form, but let us for instance assume that it is given by:

\begin{eqnarray}
U \highlight{\left[\frac{(\ket{0}+\ket{1})^3}{\sqrt{8}}\ket{\psi_0}\right]} &=&  \frac{1}{\sqrt{2}}\ket{A_1}\ket{\psi_0} +  \frac{1}{\sqrt{2}}\ket{A_2}\ket{\psi_1}\label{Ex1}\\
U \highlight{\left[\frac{(\ket{0}+\ket{1})^3}{\sqrt{8}}\ket{\psi_1} \right]}&=&  \frac{1}{\sqrt{2}}\ket{A_3}\ket{\psi_0} +  \frac{1}{\sqrt{2}}\ket{A_4}\ket{\psi_1}\label{Ex2}
\end{eqnarray}
where $\{\ket{\psi_0},\ket{\psi_1}\}$ is an orthonormal basis for $H_O$ (we have chosen a particular $U$ for which the Schmidt decomposition of the mapped states can be done in the same basis of $H_O$), and where
\begin{eqnarray}
\ket{A_1} &=&\frac{\ket{000}+\ket{001}+\ket{010}+\ket{011}}{\sqrt{4}}\\
\ket{A_2} &=& \frac{\ket{100}+\ket{101}+\ket{110}+\ket{111}}{\sqrt{4}}\\
\ket{A_3} &=& \frac{\ket{000}-\ket{001}+\ket{010}-\ket{011}}{\sqrt{4}}\\
\ket{A_4} &=& \frac{\ket{100}-\ket{101}+\ket{110}-\ket{111}}{\sqrt{4}} 
\end{eqnarray}

Suppose that the observer is initially in the state $\ket{\psi_0}$ and performs successive measurements on 3-qubit streams. We see that the reduced density matrix for each stream will be a linear combination of density matrices $\rho_S^1 = \frac{1}{2}\left(\ket{A_1}\bra{A_1}+\ket{A_2}\bra{A_2}\right)$ and $\rho_S^2 = \frac{1}{2}\left(\ket{A_3}\bra{A_3}+\ket{A_4}\bra{A_4}\right)$. If we measure the outgoing 3-qubit streams in the canonical basis $\{\ket{a b c}\}_{a, b, c = 0,1}$, then for each stream, the probability to observe $\ket{a b c}$ is 1/8 for any combination of $a$, $b$ and $c$. Therefore, in this example, measurement in the canonical basis is not sufficient to show that the 3-qubit system is not in the fully mixed state.

Now, we also note that the states $\ket{A_i}$ are all orthogonal to the following states in $H_S$:
\begin{eqnarray}
\ket{B_1} &=&\frac{\ket{000} - \ket{010}}{\sqrt{2}} = \ket{0}\ket{-}\ket{0}\\
\ket{B_2} &=&\frac{\ket{001} - \ket{011}}{\sqrt{2}} = \ket{0}\ket{-}\ket{1}\\
\ket{B_3} &=&\frac{\ket{100} - \ket{110}}{\sqrt{2}} = \ket{1}\ket{-}\ket{0}\\
\ket{B_4} &=&\frac{\ket{101} - \ket{111}}{\sqrt{2}} = \ket{1}\ket{-}\ket{1}\\
\end{eqnarray}
where $\ket{-}=(\ket{0}-\ket{1})/\sqrt{2}$. The successive interaction of the observer and the $m$ streams of qubits will keep the state of the streams orthogonal to the subspace spanned by the states $\ket{B_i}$. Therefore, measurement of the streams of 3 qubits in the basis $\{\ket{A_i},\ket{B_i}\}_i$ will lead to zero observation of the states $\ket{B_i}$ and will thus reveal that the 3-qubit streams are obviously not in the fully mixed state.

In general, the transformation given by the equations (\ref{Ex1},\ref{Ex2}) is not known, but as we have many streams of qubits sharing the same density matrix, one can find an observation basis for which the measurement outcome is not compatible with a fully mixed state.

\section{Discussion}
We have shown that Everett's theory and the Copenhagen theory give different predictions when the dimension of the observed $N$-qubit system is larger than $D^2$ where $D$ is the ``dimension'' of the observer. If we were to express the dimension of the observer in qubits, this would give $d=\log_2 D$ so our result states that it is sufficient that $N >  2 d$ to test Everett's theory against the Copenhagen theory. Expressed in loose terms, Everett's predictions and Copenhagen's diverge when the observer is performing a measurement on a system whose information content is twice the information content that he can store.

Another important point to note is that the condition $N > 2d$ is a sufficient condition and not a necessary one. An experiment for a given observer could show that the reduced density matrix of the observed system is different from the fully mixed state for $N$ well below $2d$. This is because the inner working of the observer is not known and it may turn out that every dimensionality describing the observer is not optimally coupled with the information conveyed by the observed system.  \highlight{In particular, Chao. R et al.~\cite{Chao} have shown that a system of $d$ qubits for which an operation on one qubit can slightly affect the other qubits can be described in a space of dimension polynomial in $d$, and not $2^d$ as we would expect from a system of $d$ fully independent qubits.} Experiment may show that the density matrix of the observed $N$-qubit stream may diverge from the fully mixed one for $N$ being a tiny fraction of $d$. 

A statistically significant evidence that the observed $N$-qubit system is not in a fully mixed state would be a sufficient proof that the observer does not ``collapse the wave function'' as postulated by the Copenhagen theory.

\highlight{
\section{Acknowledgements}
The author would like to thank Artur Ekert and Lev Vaidman for their comments, and in particular to David Deutsch for the helpful discussions and also for pointing out the work by R. Chao et al. on overlapping qubits.
}

\end{document}